\documentclass[12pt,preprint]{aastex}
\usepackage{emulateapj5}

\slugcomment{
submitted to ApJ
}

\shorttitle{NIR colors of GRB afterglows and reionization}
\shortauthors{Inoue, Yamazaki, and Nakamura}

\begin{document}

\title{Near-Infrared Colors of Gamma-Ray Burst Afterglows 
and Cosmic Reionization History}
\author{
Akio K.~Inoue\altaffilmark{1}, Ryo~Yamazaki\altaffilmark{1}, 
and Takashi~Nakamura
}
\affil{Department of Physics, Kyoto University, Kyoto 606-8502, Japan}
\email{
akinoue@tap.scphys.kyoto-u.ac.jp,
yamazaki@tap.scphys.kyoto-u.ac.jp, 
takashi@tap.scphys.kyoto-u.ac.jp
}

\altaffiltext{1}{Research Fellow of Japan Society for the Promotion of
Science.}

\begin{abstract}
Using the near-infrared (NIR) observations of the afterglows of the
 high redshift ($5 \la z \la 25$) gamma-ray bursts (GRBs) that will be
 detected by the {\it Swift} satellite, we discuss a way to study the
 cosmic reionization history.
In principle the details of the cosmic reionization history are well
 imprinted in the NIR spectra of the GRB afterglows. 
However the spectroscopy with a space telescope is required to obtain
 such an information for a very high redshift ($z\ga 15$) unless the
 neutral fraction of the high-$z$ universe is less than $10^{-6}$. 
The broad-band photometry has the higher sensitivity than that of the
 spectroscopy, so that the NIR photometric follow-up of the GRB
 afterglows is very promising to examine the cosmic reionization
 history. 
A few minutes exposure with a 8-m class ground-based telescope of the
 afterglow of the high-$z$ GRBs will reveal how many times the
 reionization occurred in the universe. 
\end{abstract}


\section{INTRODUCTION}

The evolutionary study of the intergalactic medium (IGM) is now one
of the most active fields in astrophysics. Especially, the cosmic
reionization history of the IGM attracts many researchers in recent
years (see \citealt{loe01} for a review). Gunn--Peterson trough
shortward of the Ly$\alpha$ resonance \citep{gn65} in the spectrum of
quasars with redshift $z\sim 6$ indicates that the end of the
reionization epoch is $z\sim 6$ \citep{bec01,whi03}. On the other hand,
the recent observation of the polarization of the cosmic microwave
background (CMB) by {\it WMAP} suggests that the beginning of the
reionization is $z\sim 20$ \citep{spe03,kog03}. One might think that
there is a discrepancy between both observations.

The detailed simulation of the six-dimensional radiative transfer shows
that the reionization process proceeds slowly in an inhomogeneous
universe \citep[e.g.,][]{nak01}. If the reionization began at $z\sim
20$, the neutral fraction in the universe decreased gradually, and the
universe was ionized almost completely at $z\sim 6$, the apparent
discrepancy can be resolved \citep{wyi03a,hai03,cia03,chi03,sok03,onk03}.   
The scenario that the initial partially ionized epoch was simply
followed by the complete ionization can explain the large Thomson
scattering opacity observed from the CMB polarization and also is
consistent with the previous picture in which the end of the
reionization is $z\sim 6$. 

\cite{cen03a,cen03b} and \cite{wyi03b} recently proposed a new scenario
of the reionization; the universe was reionized twice. 
Even in an inhomogeneous universe, enough strong intensity of the
ultra-violet (UV) background radiation can ionize the universe quickly.
At $z\sim 20$, the first reionization was made by the Population III
(Pop III) stars (metal-free stars) with a top-heavy initial mass
function (IMF) which yields a much higher UV emissivity than that of
normal Pop I and II stars.  Then, the universe
was partially recombined at $z\sim 15$ when the transition from the Pop
III to II was occurred and the UV emissivity was suddenly suppressed
because of the different IMF. Finally, the UV photons from the Pop II
stars increased gradually and ionized the universe again at $z\sim 6$.  

Cen's scenario can also resolve the discrepancy, but it is different
from others in the history of the cosmic reionization; 
the first complete ionization was followed by the partially ionized
epoch and then the second complete ionization.
We should assess whether Cen's scenario is favored by using 
other observations. In this paper, an assessment by using
the afterglows of the gamma-ray bursts (GRBs) is discussed. 

The usefulness of GRBs to investigate the high-$z$ universe is pointed
out by many authors \citep[e.g.,][]{lamb00,ciardi00,bar03}.
It is strongly suggested that the long duration GRBs arise from the
collapse of a massive star
\citep{galama98,uemura03,hjorth03,price03,sta03,mat03}.  
Hence, the GRBs can occur at very high-$z$ 
once the massive stars are formed. For example, the {\it Swift} 
satellite\footnote{http://swift.gsfc.nasa.gov/} is expected to detect
$\sim 10$ GRBs per year occurring at $z\ga 10$ \citep{lamb00}. 
Furthermore, if we fix an observing time from the prompt emission, the
observed afterglow flux does not become so faint even for the extremely 
high-$z$ GRBs because the earlier phase of the afterglow 
is observed in the cosmological rest frame \citep{ciardi00}.  

In this paper, we show that the near-infrared (NIR) photometric
follow-ups of the GRB afterglows are very useful to investigate how
many times the cosmic reionization occurred. Although many techniques to
prove the reionization history have been proposed so far, for example,
hydrogen 21 cm line tomography \citep{mad97,car02,fur02}, Ly$\alpha$
damping wing measurement \citep{mir98,bar03}, metal absorption lines
\citep{oh02,fur03}, CMB polarization anisotropy \citep{hai03}, and
dispersion measure in GRB radio afterglows \citep{iok03,ino03}, 
the NIR photometric colors may be the most promising technique using the
{\it current} facilities.
 
In order to investigate the reionization history using our
photometric method, we need to determine 
redshifts of GRBs by other ways, e.g., detections of iron lines in
X-ray afterglow spectra \citep{meszaros03} and of the Ly$\alpha$ break in
NIR spectra, or empirical methods by using only the $\gamma$-ray data
\citep{fen00,norris00,in01,ama02,att03,mur03,yon03}. 
Even if redshifts of GRBs are unknown prior to follow-up observations,
it is worth performing NIR photometry as early phase as possible 
since $\sim 10$\% of GRBs are expected 
to be located at $z\ga 10$ \citep{bro02}. 
In practice, such follow-up observations against every GRB are possible.
Even after the early NIR follow-up, we will be able to
determine the redshifts. 
In this paper, we show that the near-infrared (NIR) photometric
follow-ups of the GRB afterglows provide a significantly
important information on the cosmic reionization history.

The structure of this paper is as follows: 
we start from a brief summary of the IGM opacity by the
neutral hydrogen in section 2. In section 3, the possible scenario of a
very low neutral fraction at very high-$z$ is discussed. Then, we
examine the NIR spectra and colors of the GRB afterglows in section 4
and 5, respectively. Finally, we discuss a way to obtain the cosmic
reionization history and advantage/disadvantage of our method 
in section 6. 

We adopt a standard set of the $\Lambda$CDM cosmology throughout the 
paper: $H_0=70$ km s$^{-1}$ Mpc$^{-1}$, $\Omega_{\rm M}=0.3$, 
$\Omega_\Lambda=0.7$, and $\Omega_{\rm b}=0.04$.

\section{IGM OPACITY}

Suppose an observer at $z=0$ who observes a source with $z=z_{\rm S}$ at 
the observer's frequency $\nu_0$. The radiation from the source is
absorbed by the Lyman series lines and the photoionization of the
intervening neutral hydrogen \citep{gn65}. The hydrogen cross section at
the rest frame frequency $\nu$ is 
\begin{equation}
 \sigma_{\rm HI}(\nu)=\sigma_{\rm LC}(\nu) + \sum_i \sigma_i(\nu)\,,
\end{equation}
where $\sigma_{\rm LC}(\nu)$ is the cross section for the Lyman
continuum photons ($h\nu \geq$ 13.6 eV) and $\sigma_i(\nu)$ is the cross
section for the $i$-th line of the Lyman series, i.e., $i=$ Ly$\alpha$,
Ly$\beta$, Ly$\gamma$, etc. Here we consider the lines up to $i=40$.
The optical depth for the observer's frequency $\nu_0$ is given by  
\begin{equation}
 \tau_{\nu_0}(z_{\rm S})=\tau_{\nu_0}^{\rm LC}(z_{\rm S}) 
  + \sum_i \tau_{\nu_0}^i (z_{\rm S})\,,
\end{equation}
\begin{equation}
 \tau_{\nu_0}^{\rm LC}(z_{\rm S})
  =\int_0^{z_{\rm S}} \sigma_{\rm LC}(\nu_0[1+z]) 
  n_{\rm HI}(z) \frac{c\, dz}{(1+z)H(z)}\,,
\end{equation}
and
\begin{equation}
  \tau_{\nu_0}^i (z_{\rm S})=\int_0^{z_{\rm S}} \sigma_i(\nu_0[1+z]) 
  n_{\rm HI}(z) \frac{c\, dz}{(1+z)H(z)}\,,
\end{equation}
where $n_{\rm HI}(z)$ is the number density of the neutral hydrogen at
the redshift $z$, $H(z)$ is the Hubble constant at the redshift $z$, and
$c$ is the light speed. 

The cross section for the Lyman continuum photons is given by 
$\sigma_{\rm LC}(\nu) = \sigma_{\rm L} (\nu/\nu_{\rm L})^{-3}$, 
where $\sigma_{\rm L}=6.30\times 10^{-18}$ cm$^2$ and the Lyman limit
frequency $\nu_{\rm L}=3.29\times 10^{15}$ Hz \citep{ost89}. 
Thus, we obtain 
\begin{equation}
  \tau_{\nu_0}^{\rm LC}(z_{\rm S})
   = \sigma_{\rm L} \left(\frac{\nu_{\rm L}}{\nu_0}\right)^3 n_{\rm H,0} c 
   \int_{z_{\rm min}}^{z_{\rm S}} \frac{x_{\rm HI}(z) dz}{(1+z)H(z)}\,,
\end{equation}
where $x_{\rm HI}(z)$ is the neutral fraction at the redshift $z$ and
$z_{\rm min}={\rm max}[0,(\nu_{\rm L}/\nu_0)-1]$. The neutral fraction
$x_{\rm HI}(z)$ is defined as $n_{\rm HI}(z)/n_{\rm H}(z)$, 
where $n_{\rm H}(z)=n_{\rm H,0}(1+z)^3$ is the cosmic mean number
density of hydrogen atom at the redshift $z$ with $n_{\rm H,0}$ being
the present one. 
The effect of the inhomogeneity of the universe can be expressed in the
definition of $x_{\rm HI}$ (see section 3 below). 
Since we need the condition $z_{\rm min} < z_{\rm S}$ to integrate
Eq. (5), $\tau_{\nu_0}^{\rm LC}(z_{\rm S})=0$ when  
$\nu_0 \leq \nu_{\rm L}/(1+z_{\rm S})$.

The line cross section $\sigma_i(\nu)$ has a very sharp
peak at the line center frequency. The width is well characterized by
the Doppler width, $\Delta \nu_{\rm D}$. That is, 
$\sigma_i(\nu) \simeq \sigma_i$ for $|\nu-\nu_i| \leq \Delta \nu_{\rm D}$, 
and 
$\sigma_i(\nu) \simeq 0$ for $|\nu-\nu_i| > \Delta \nu_{\rm D}$, 
where $\sigma_i$ and $\nu_i$ are the cross section and the frequency of
the $i$-th line center, respectively. Since  $\nu=\nu_0 (1+z)$, we
integrate the rhs of equation (4) over a narrow redshift width of 
$\Delta (1+z) = 2 (\Delta \nu_{\rm D}/\nu_i)(1+z_i)$ 
around $1+z_i = \nu_i/\nu_0$. Therefore, we obtain 
\begin{equation}
 \tau_{\nu_0}^i (z_{\rm S}) \simeq 2
  \left[\frac{\sqrt{\pi} e^2 f_i}{m_e c \nu_i}\right]
  \left[\frac{c n_{\rm HI}(z_i)}{H(z_i)}\right]\,,
\end{equation}
where we have substituted 
$\sigma_i=\sqrt{\pi}e^2f_i/m_e c \Delta \nu_{\rm D}$, 
in which $e$ is the electric charge, $f_i$ is the absorption oscillator
strength of the $i$-th line, and $m_e$ is the electron mass.  
The values of $f_i$ and $\nu_i$ are taken from \cite{wie66}.
This time also  we are  restricted by $z_i \leq z_{\rm S}$. Thus, the
above equation is valid only when $\nu_0 \geq \nu_i/(1+z_{\rm S})$,
otherwise $\tau_{\nu_0}^i (z_{\rm S})=0$.

For the Ly$\alpha$ line, the opacity becomes 
\begin{equation}
 \tau^{\rm Ly\alpha} \simeq 2.6 \times 10^6 x_{\rm HI}(z) 
  \left(\frac{1+z}{20}\right)^{3/2}\,,
\end{equation}
where we approximate $H(z)\approx H_0 \Omega_{\rm M}^{1/2} (1+z)^{3/2}$ 
and adopt $n_{\rm HI}(z)=x_{\rm HI}(z)n_{\rm H,0}(1+z)^3$.
This is the Gunn-Peterson optical depth \citep{gn65,pee93}. 
{}From this equation, we realize that if $x_{\rm HI} \ga 10^{-6}$, the
IGM is opaque for the photon bluer than the source Ly$\alpha$ line. 
Conversely, we can estimate $x_{\rm HI}$ if $\tau^{\rm Ly\alpha}$ is
determined observationally (e.g., \citealt{bec01,whi03}).

\section{NEUTRAL FRACTION AND STAR FORMATION RATE}

In the following sections, we will assume a neutral fraction of
hydrogen, $x_{\rm HI} \sim 10^{-6}$ at $z\sim 20$ for calculations of
spectra and colors of GRB afterglows. Before such calculations, we show,
here, that $x_{\rm HI}\sim 10^{-6}$ is possible at a very high-$z$
universe.\footnote{In Figure 9 of \cite{cen03a}, the neutral fraction
decreases up to $10^{-4}$. However, it is an artificial limit in his
calculation (Cen 2003, private communication).}

In general, the neutral fraction, $x_{\rm HI}(z)$, is spatially
variable. However, we discuss only its mean value for
simplicity.  The averaged $x_{\rm HI}$ is determined by the UV 
intensity of the background radiation and the inhomogeneity of the
hydrogen number density at the redshift $z$.  Even at very
high-$z$ universe, the recombination time-scale is much less than
the Hubble time-scale. Thus, we assume the ionization equilibrium. In
that case, the mean neutral fraction for a highly ionized
medium (i.e. $x_{\rm HI} \ll 1$) is given by 
$x_{\rm HI} \approx C n_{\rm H} \alpha/\Gamma_{\rm HI}$, 
where $C=\langle n_{\rm H}^2 \rangle/\langle n_{\rm H} \rangle^2$ 
is the clumping factor, $n_{\rm H}$ is the number density of hydrogen
nuclei, $\alpha$ is the recombination coefficient, and  
$\Gamma_{\rm HI}=\int \sigma_{\rm LC}(\nu) c n_\nu d\nu$ is the HI
photoionization rate where $n_\nu$ is the photon number density per
frequency. Therefore, the UV background intensity is required to
estimate $x_{\rm HI}$ even if we know the clumping factor from the model
of the cosmological structure formation. 

How much photons are required to control the hydrogen neutral fraction
in a very low level? Approximately, the HI photoionization rate is
estimated to be $\Gamma_{\rm HI} \sim \sigma_{\rm L}c n_{\rm ion}$ with
$n_{\rm ion}$ being the number density of ionizing photons. Hence, we
find 
\begin{equation}
 \frac{n_{\rm ion}}{n_{\rm H}} 
  \sim \frac{C\alpha}{\sigma_{\rm L}cx_{\rm HI}}
  \sim C \left(\frac{10^{-6}}{x_{\rm HI}}\right)\,,
\end{equation}
where the case B recombination rate 
$\alpha=2.73\times10^{-13}$ cm$^3$ s$^{-1}$ is adopted \citep{ost89}.
Now, we are interested in a very high redshift universe ($z\sim20$). 
In there, the clumping factor is an order of unity. Therefore, we find
that only one photon per a hydrogen nuclei is sufficient to keep 
$x_{\rm HI} \sim 10^{-6}$.

Then, we examine how much stars are required to maintain the photon
density. Since a proton recombines with an
electron in a certain time-scale, continuous supply of ionizing photons
is needed to keep $n_{\rm ion} \sim n_{\rm H}$. Since the
recombination time-scale is $\sim 1/n_{\rm H}\alpha$ 
for $x_{\rm HI} \ll 1$, the required photon 
emissivity per unit volume is estimated to be 
$\epsilon_{\rm ion}\sim n_{\rm ion}n_{\rm H}\alpha
\sim Cn_{\rm H}^2\alpha^2/x_{\rm HI}\sigma_{\rm L}c$.
On the other hand, the emissivity is given by 
$\epsilon_{\rm ion}=f_{\rm esc}\epsilon_{\rm LC}\rho_{\rm SFR}(1+z)^3$,
where $f_{\rm esc}$ is the escape fraction of Lyman continuum from
primordial galaxies, $\epsilon_{\rm LC}$ is the Lyman continuum photon
emissivity per unit stellar mass, and $\rho_{\rm SFR}$ is
the star formation density per unit time per unit comoving volume.
Therefore, the required star formation density is 
\begin{eqnarray}
 \frac{\rho_{\rm SFR}}{M_\sun\,{\rm yr^{-1}\,Mpc^{-3}}}
  &\sim& 0.1C \left(\frac{0.1}{f_{\rm esc}}\right)
  \left(\frac{10^{61}\,{\rm ph}\,{M_\sun}^{-1}}{\epsilon_{\rm LC}}\right)
  \nonumber \\ && 
  \left(\frac{10^{-6}}{x_{\rm HI}}\right)
  \left(\frac{1+z}{20}\right)^3\,.
\end{eqnarray}

Although the photon emissivity at high-$z$ is very uncertain because we
don't know the stellar mass distribution, we estimate the emissivity by
Starburst 99 model
\citep{lei99}\footnote{http://www.stsci.edu/science/starburst99/} by
assuming the Salpeter mass function with various mass ranges. The
estimated values of $\epsilon_{\rm LC}$ are summarized in Table 1. 
The Pop III stars are likely to have a top-heavy mass function 
\citep[e.g.,][]{nu01}. Hence, the case of 10--100 $M_\sun$ in Table 1
may be suitable. In this case, the required star formation density to
maintain $x_{\rm HI}\sim10^{-6}$ is $\sim 0.05$ $M_\sun$ yr$^{-1}$
Mpc$^{-3}$ (comoving) when $f_{\rm esc}=0.1$ and $C=1$. 

Is this star formation density possible? 
The latest observations suggest that the star
formation density retains a level of $\sim 0.1$ $M_\sun$ yr$^{-1}$
Mpc$^{-3}$ from $z\sim 1$ to $z\sim 6$ \citep{gia03}. Such a level of
star formation may be kept toward more high-$z$ universe. Moreover, a
semianalytic model shows 0.01--0.1 $M_\sun$ yr$^{-1}$ Mpc$^{-3}$ at
$z\sim 20$ depending on the assumed star formation efficiency
\citep{som03}. Thus, we can sufficiently expect 
$\sim 0.05$ $M_\sun$ yr$^{-1}$ Mpc$^{-3}$, and then, 
$x_{\rm HI}\sim10^{-6}$ at $z\sim20$. In addition, we note that 
the escape fraction may be much larger than 0.1 assumed above 
if the Pop III stars are formed in low-mass halos.

\section{NEAR INFRARED SPECTRA}

Let us discuss the observed spectra of the GRB afterglows in
the NIR bands. To do so, an afterglow spectral model is required. We 
adopt a simple afterglow model; the synchrotron radiation from the
relativistic shock \citep{sar98}. More specifically, we adopt equations
(1)--(5) in \cite{ciardi00} who take into account the effect of the
cosmological redshift. The adopted parameters are the magnetic energy
fraction of $\epsilon_{\rm B}=0.1$, the electron energy fraction of
$\epsilon_{\rm e}=0.2$, the spherical shock energy of $E=10^{52}$ erg,
the ambient gas number density of $n=10$ cm$^{-3}$, and the power-law
index of the electron energy distribution $p=2.5$. 

First we consider the observed afterglow spectra in the
hypothetical perfectly neutral universe for comparison. 
In Figure 1, we show the expected afterglow spectra in the neutral
universe observed 1 hour after the burst in the observer's
frame. Due to the Ly$\alpha$ line absorption, the continuum bluer
than the Ly$\alpha$ line in the source frame (the observed wavelength
0.1216[$1+z_{\rm S}$] \micron) is completely damped. Thus,
we can find the Ly$\alpha$ break clearly. From the observed wavelength
of the Ly$\alpha$ break, we can determine the redshift of the GRBs.
If we observe the afterglow
through a filter, the radiation from the source with the redshift
higher than the characteristic redshift of the filter cannot be
detected because of the Ly$\alpha$ break. For example, the effect starts
from $z_{\rm S}\simeq 8$ for the $J$-band and the source with 
$z_{\rm S}\ga 11$ cannot be seen through the filter, 
i.e. $J$ {\it drop-out}. These characteristic redshifts are summarized
in Table 2. However, we have a chance to see the source
beyond the drop-out redshift if the universe is highly ionized as we
will show later. 

Next we examine what is observed if the very high-$z$ universe is
ionized completely as proposed by \cite{cen03a,cen03b}. Let us set the
neutral fraction, $x_{\rm HI}$, of the universe in the redshift range
$15\leq z < 20$ to be very small and $x_{\rm HI} \sim 1$ for 
$z < 15$ and $z \geq 20$. That is, we assume that the Pop III stars
ionized the universe at $z=20$ and the sudden change of the IMF due to
the transition from the Pop III to II was occurred at $z=15$. 
The neutral fraction is determined by the background UV intensity
produced by the Pop III stars. However, the intensity is
quite uncertain, so that we choose two cases as $x_{\rm HI}=10^{-6}$ and
$10^{-7}$ for instance (see also Fig.5 in section 5). 
We will present a way to constrain $x_{\rm HI}$
from the observations later. 

In Figure 2, we show the expected spectra of the GRB
afterglows. The solid, short-dashed, long-dashed, dot-dashed, and
dotted curves are the expected afterglow spectra of the GRB at 
$z_{\rm S}=22$, 20, 18, 15, and 13, respectively. 
The observing time is assumed to be 1 hour after the
prompt emission in the observer's frame. We find the clear 
Ly$\alpha$ breaks in the spectra. Since $x_{\rm HI}=10^{-6}$ (or
$10^{-7}$) at $15\leq z < 20$ is assumed in the panel (a) (or (b)), 
the IGM opacity is an order of unity (or 0.1) in the redshift range (see
eq.[7]).  Thus, the continuum bluer than the Ly$\alpha$ break of
the GRBs with $z_S > 15$ still remains of the order of $10$ $\mu$Jy (or
100 $\mu$Jy) in $\sim 2$--2.5 \micron, i.e.,in  $K$-band 
(see the thin solid curve indicated as $K$; we
adopt the filter system of \citealt{bes88}). 
The spectral break at 1.94(=0.1216[1+15]) \micron\ is due to the neutral
hydrogen below $z=15$. Thus, the continuum less than the break wavelength in
the observer's frame from the $z_{\rm S}\geq 15$ source is
completely extinguished. The spectra of the 
source with $z_{\rm S} < 15$ are the same as those shown in Figure 1.
For $z_{\rm S}=22$ case, the spectrum shows the second break at
2.36(=0.1026[1+22]) \micron. This is the Ly$\beta$ break due to the
neutral hydrogen near the GRB. Here we have assumed $x_{\rm HI}\sim 1$ 
for $z\geq 20$.

The structure corresponding to the reionization history appears in the
spectra of the GRB afterglows as shown in Figure 2 
(see also \citealt{hai99}). If
we could detect the continuum rising at 2.55(=0.1216[1+20]) \micron\
in the spectrum of the GRB with $z_{\rm S}=22$, we would
find the starting epoch of the first reionization as $z=20$. 
On the other hand, the end of the first
reionization, in other words, the transition epoch from the Pop III to
II is realized from the spectral break at 1.94(=0.1216[1+15])
\micron. 

The suitable band to determine the ionization history depends on the
redshift of the reionization epoch. 
From Figure 1 (see also Table 2), we find that the $I$, $J$, $H$,
and $K$-band spectroscopies are suitable for the reionization epoch at
$z\simeq 5$--7, 8--11, 11--14, and 15--20, respectively.
In any case, observations to detect the spectral signatures in the NIR
afterglow spectra of the GRBs are strongly encouraged. 
If $x_{\rm HI}$ is smaller than $10^{-6}$, we can clearly see the
difference from Fig.~1 and confirm the double reionization.  

In the above discussion, we assumed $x_{\rm HI}=10^{-6}$ and $10^{-7}$. 
Let us argue what will happen for different values of  $x_{\rm HI}$.
If $x_{\rm HI}\ga 10^{-6}$, the remaining flux decreases 
exponentially because the IGM opacity becomes
much larger than unity; for example, the flux is 
about 1 nJy for $x_{\rm HI}=6\times10^{-6}$, 
and about 1 pJy for $x_{\rm HI}=10^{-5}$.  
It is quite difficult to detect such a low-level flux.

\section{NEAR INFRARED COLORS}

The photometric observations are available easier than the
spectroscopy. We examine the expected apparent NIR colors of
the GRB afterglows. Although we can discuss the apparent magnitude of
the afterglows in one photometric filter, 
their dispersion is very large because the luminosity of
the afterglows depends on many uncertain parameters such as the jet
opening angle, the ambient matter density, the magnetic energy fraction 
and the relativistic electron energy fraction. 
On the other hand, the dispersion of the apparent colors
can be quite small because the color does not depend on the absolute
luminosity but only the spectral shape which does not change
significantly in the observed NIR bands.   

The apparent magnitude\footnote{All magnitudes in this paper are the
Vega system.} in a filter band denoted as $i$ is defined by 
\begin{equation}
 m_i = -2.5 \log F_i/F_{i,0}\,,
\end{equation}
where $F_{i,0}$ is the zero point flux density of the filter and $F_i$ is
the mean flux density through the filter which is  
\begin{equation}
 F_i = \frac{\int T_{i,\nu}f_\nu e^{-\tau_\nu} d\nu}
  {\int T_{i,\nu} d\nu}\,,
\end{equation}
where $T_{i,\nu}$ is the transmission efficiency of the filter and 
$f_\nu e^{-\tau_\nu}$ is the flux density entering the filter. If there
is no intervening absorber between the source and the telescope, 
$\tau_\nu=0$. 

The observed color between two filter bands, $i$ and $j$ (the filter $i$
is bluer than the filter $j$), is given by 
\begin{equation}
 m_i - m_j = (m_i - m_j)^{\rm int} + (\Delta m_i - \Delta m_j)\,,
\end{equation}
where $(m_i - m_j)^{\rm int}$ is the intrinsic color of the source, 
and $\Delta m_i$ and $\Delta m_j$ are the absorption amounts 
in the filters $i$ and $j$, respectively. 
When we consider the NIR filter bands and high-$z$ GRBs (for example
$z=15$), the intrinsic afterglow spectrum is predicted to be
proportional to $\nu^{-1/2}$ from $\sim 1$ minute to several hours after
the burst occurrence and proportional to $\nu^{-p/2}$ for later time
in the standard afterglow model \citep{sar98,ciardi00}. 
Other parameters adopted are described in the
beginning of the section 3. In Table 3, we tabulate the intrinsic colors
of the sources for the two cases of the spectral shape.  

The absorption amount in the filter $i$ is  
\begin{equation}
 \Delta m_i \equiv m_i - m_i^{\rm int}
  = -2.5 \log \frac{\int T_{i,\nu}f_\nu e^{-\tau_\nu} d\nu}
  {\int T_{i,\nu} f_\nu d\nu}\,,
\end{equation}
where $m_i^{\rm int} = m_i(\tau_\nu=0)$ is the intrinsic (no absorption)
apparent magnitude. In a certain band width $\Delta \nu$, 
the difference in the optical depth 
$\Delta \tau$ is estimated as 
$| \Delta \tau/\tau | = 3/2 (1+z)^{-1} | \Delta (1+z) | \sim (1+z)^{-1}
| \Delta \nu/\nu |$ from equation (7).
Since  $\Delta \nu$ of the filter transmission is smaller than
the effective frequency of the filter, 
i.e. $\Delta \nu/\nu < 1$, and also $z\ga 5$,  
$\Delta \tau/\tau \ll 1$, that is, the term $e^{-\tau_\nu}$ in
the integral of the numerator in equation (13) can be regarded as
almost constant. Hence, we obtain approximately 
$\Delta m_i \approx 1.086 \tau_{\rm eff}$, where $\tau_{\rm eff}$ is the
effective IGM opacity in the filter $i$.

Now we can estimate the observed color by equation (12) if $\Delta m_i$
is known. 
To know $\Delta m_i$ is equivalent to know the IGM effective
opacity $\tau_{\rm eff}$. This opacity is one between the redshift at
which the Ly$\alpha$ break comes into the filter band width  
($z_{\rm Ly\alpha,in}^i$, see Table 2) 
and the source redshift ($z_{\rm S}$) because 
the neutral hydrogen in $z<z_{\rm Ly\alpha,in}^i$ cannot absorb the
photons passing through the filter $i$. We note that the neutral hydrogen
lying beyond the redshift at which the Ly$\alpha$ break goes out of the
filter band width ($z_{\rm Ly\alpha,out}^i$, see Table 2) absorb the
photons through the filter $i$ because of the higher-order Lyman series
lines like Ly$\beta$, Ly$\gamma$, etc., and the photoionization process.
As a result, $\tau_{\rm eff}$ is determined by the neutral fraction
$x_{\rm HI}$ in the redshift range 
$z_{\rm Ly\alpha,in}^i \leq z \leq z_{\rm S}$.
Since $x_{\rm HI}$ at high-$z$ is uncertain, we assume that 
$x_{\rm HI}$ is constant in the above range for simplicity.   
The real $x_{\rm HI}$ might vary significantly in the redshift range, 
so that the assumed $x_{\rm HI}$ should be regarded as an effective
mean value including such a variation (hereafter $x_{\rm HI}^{\rm eff}$).

In Figures 3a--3d, we show $\Delta m_i$ for the $I$, $J$, $H$, and
$K$-bands as a function of the source redshift. 
The continuum in the observer's $L$-band is not absorbed 
by the IGM neutral hydrogen at all when the source
redshift is less than about 25. Although we assumed that 
the spectral shape is proportional to $\nu^{-1/2}$ in the panels, 
the results are much robust for the change of the spectral shape 
as noted above.  The solid curves in these panels are
loci of $\Delta m_i$ for a given $x_{\rm HI}^{\rm eff}$ 
as a function of the source redshift $z_{\rm S}$. 
For example, since $z_{\rm Ly\alpha,in}^I=5$ for $I$-band, the IGM
absorption in $I$-band is about 5 mag for the source at $z_{\rm S}=7$ if 
the effective neutral fraction $x_{\rm HI}^{\rm eff}$ in the redshift
range $5 \leq z \leq 7$ is $10^{-5}$.

Two dotted vertical lines in each panel of Figure 3 indicate the
redshifts when the Ly$\alpha$ break enters and goes out of the
each band width ($z_{\rm Ly\alpha,in}^i$ and $z_{\rm Ly\alpha,out}^i$,
respectively). As seen in Figure 1, if the IGM is significantly neutral,
the source with $z_{\rm S} > z_{\rm Ly\alpha,out}^i$ cannot be
detected through the filter $i$ (i.e. {\it drop-out}). However, we can
detect such a source if the universe is highly ionized.
This is because the continuum below the Ly$\alpha$ break remains as
shown in Figure 2.

Suppose we observe a source with $z_{\rm S}>z_{\rm Ly\alpha,out}^i$
through the $i$ and $j$ filters and assume that the radiation through
$j$ filter is not affected by any intervening absorption 
(i.e. $\Delta m_j=0$). From equation (12), 
the expected magnitude through the $i$ filter is 
\begin{equation}
 m_i = m_j + (m_i-m_j)^{\rm int} + \Delta m_i\,.
\end{equation}
For example, we consider the case of $i=I$, $j=L$, and $z_{\rm S}=7$.
We find that the apparent $L$ magnitude of the afterglow of the GRB at 
$z_{\rm S}=7$ for 1 hour after the prompt burst is about 14 mag from
Figure 4. Thus, the apparent $I$ magnitude is expected to be 22 mag
because the intrinsic $I-L=3.1$ mag for 1 hour (i.e. the case 
$\propto \nu^{-1/2}$ in Table 3) and $\Delta I\sim 5$ mag if 
$x_{\rm HI}^{\rm eff}=10^{-5}$ in the redshift range $5 \leq z \leq 7$. 
We can reach 5-$\sigma$ detection of the source with $I=22$ mag by only 
three minutes exposure with a 8-m class telescope. Interestingly the
assumed $x_{\rm HI}^{\rm eff}$ is similar to the value reported by
\cite{whi03} from the Gunn--Peterson trough in the spectra of $z>6$
quasars.   

Similar argument can be done for other bands. Therefore, in general,  
the detection of the source with 
$z_{\rm S}>z_{\rm Ly\alpha,out}^i$ through the filter $i$ is the very
good evidence that the universe in the redshift range 
$z_{\rm Ly\alpha,in}^i \la z \la z_{\rm S}$ is highly ionized,  
i.e. $x_{\rm HI}^{\rm eff} \ll 1$ in that redshift range.
Conversely, the detection enables us to estimate $\tau_{\rm eff}$ and 
$x_{\rm HI}^{\rm eff}$.
Finally, we note here that Figure 3 is also useful for any other sources
(e.g., QSOs) because $\Delta m_i$ is almost independent of the source
spectrum.

\section{DISCUSSIONS}

\subsection{Was the universe reionized twice?}

Here we discuss how we can confirm or refute Cen's scenario: the
universe was reionized twice.
In the scenario, the first complete ionization at $z\sim 20$ is
followed by the partially ionized epoch at $z\sim 10$. Therefore, we
should check whether the neutral fraction at $z\sim 20$ is very low or
not and whether the fraction at $z\sim 10$ is almost unity or not.
To do so, the best observation is the spectroscopy in the NIR
bands. As shown in  Fig. 2 of section 3, the reionization history is
imprinted in the observed continuum.
However, the sensitivity of the spectroscopy is in
general much less than that of the photometric observations. 
Hence we discuss the way using the NIR photometries.

In the previous section, we have shown that the detection of the GRB
afterglows through a filter $i$ beyond the Ly$\alpha$ drop-out redshift  
($z_{\rm Ly\alpha,out}^i$) proves the ionization of the universe around
$z_{\rm Ly\alpha,out}^i$. From the characteristic redshifts for the NIR
filters summarized in Table 2, the $I$, $J$, $H$, and $K$ filters are
the suitable to check the ionization state at $z\sim 5$--8, 8--11,
11--15, and 15--20, respectively. Thus, the null detection of the GRB
afterglows of $z_{\rm S}\sim 11$ in $J$-band indicates an  high neutral
fraction in $8 \la z \la 11$. 
On the other hand, we detect the GRB afterglows
of $z_{\rm S}\sim 20$ in $K$-band if the IGM in $15 \la z \la 20$ is
ionized.  $I$ and $H$-band surveys are also very important to
assess the reionization history of the universe.  We can
constrain the latest reionization epoch by observing the GRB afterglows
at $z\ga 6$ through $I$-band. In summary, we can examine the
reionization history by checking whether the high-$z$ GRB afterglows
drop out of the NIR filters or not.

In the rest of this subsection, we discuss what is the difference between 
Cen's scenario and others. To demonstrate the main feature, we assume two
schematic reionization histories; (1) single gradual reionization and (2)
double sudden reionizations, which are shown in Figure 5 as the solid and
dashed curves, respectively. These histories are based on two
observational constraints; (1) the neutral fraction of hydrogen  
$x_{\rm HI}\sim 10^{-5}$ at $z\sim 6$ from the Gunn-Peterson trough
found in the $z>6$ quasars spectra \citep{bec01,whi03}, and (2) the
beginning of the reionization is $z\sim 20$ from the large opacity of
the electron scattering suggested by {\it WMAP} \citep{kog03}. For the
double reionizations, we also consider different values of $x_{\rm HI}$
in the first reionization epoch.

In Figure 6, we show the expected NIR colors as a function of the source
redshift for the afterglow spectrum $f_\nu \propto \nu^{-1/2}$ case (the
observing time less than several hours, see Table 3).
We find differences between the single and double reionizations in the
$I-J$ (panel [a]) and $K-L$ (panel [d]) colors. In panel (a), the
GRB afterglows up to $z_{\rm S} \sim 8$ can be seen in both of 
the $I$ and $J$ bands for the single reionization case, whereas the $I-J$
color of $z_{\rm S} > 7.1$ afterglows diverges in the double reionization
case, i.e. the sources drop out of the $I$-band. This reflects the
difference of the increasing rate of the neutral fraction around 
$z\sim6$ in two reionization histories. The drop-out redshift in the
double reionization case is determined by 
$\lambda_{\beta}(1+z_{\rm S,drop})=\lambda_{\alpha}(1+z_{\rm reion})$, 
where $\lambda_{\alpha}$ and $\lambda_{\beta}$ are the rest-frame
wavelength of the Ly$\alpha$ and Ly$\beta$ lines, and $z_{\rm reion}$ is
the sudden reionization redshift \citep{hai99}. In our case, 
$z_{\rm reion}=6$. It is worth to noting that the GRB afterglows beyond
the drop-out redshift of the $I$-band ($z_{\rm Ly\alpha,out}^I=6.6$) 
can be seen through the filter in both of reionization histories. 

In panel (d) of Figure 6, we find a significant difference between the
two reionization scenarios. While the afterglows beyond the drop-out
redshift of the $K$-band really drop out of the filter for the single 
reionization, we can see such afterglows in the twice reionized
universe. On the other hand, the afterglows with $z_{\rm S} > 23$ drop
out of the $K$-band even in the double reionization case 
because the Ly$\beta$ break goes out of the filter
transmission width, i.e. $\lambda_{\beta}(1+z_{\rm S,drop}) = 
\lambda_{\rm max}^K [= \lambda_{\alpha}(1+z_{\rm Ly\alpha,out}^K)]$, 
where $\lambda_{\rm max}^K$ is the maximum wavelength of the $K$-band
filter. We note here that $z_{\rm reion}(=20)>z_{\rm Ly\alpha,out}^K$ in
this time, whereas $z_{\rm reion}(=6)<z_{\rm Ly\alpha,out}^I$ in the
above case.

Any difference between the single and double reionizations cannot be
found in the $J-H$ (panel [b]) and the $H-K$ (panel [c]) colors because
the neutral fractions in both cases are high enough ($\ga 10^{-4}$) to
extinguish the continuum bluer than the Ly$\alpha$ break completely.
That is, the GRB afterglows beyond the drop-out redshift cannot be seen
in the $J$ and $H$-bands for both of reionization histories.

Let us summarize how to confirm or refute Cen's scenario: 
We can conclude that the universe was reionized twice if (1) the GRB
afterglows with $z_{\rm S} > z_{\rm Ly\alpha,out}^K$ is detected in the
$K$-band and (2) the afterglows with 
$z_{\rm S}>z_{\rm Ly\alpha,out}^J$ or $z_{\rm S}>z_{\rm Ly\alpha,out}^H$ 
drop out of the $J$ or $H$-bands. However, the null detection of the 
$z_{\rm S} > z_{\rm Ly\alpha,out}^K$ GRB afterglows in the $K$-band does
not reject Cen's scenario at once. The null detection only shows the
neutral fraction at $z\sim 20$ larger than $\sim 10^{-5}$. 
In any case, the deep and prompt $K$-band photometry of the high-$z$
GRB afterglows is useful to examine the ionization state at $z\sim 20$
very much.

\subsection{Comment on Lyman break technique}

In the above discussions, we assumed that the GRB redshifts are 
known from other methods. 
Although the spectroscopy of the optical afterglow or the host galaxies
is used to determine the redshift of the low-$z$ GRBs, it may be
difficult for the high-$z$ GRBs. The Fe line in the X-ray afterglow
\citep{meszaros03} or some empirical ways using the
$\gamma$-ray data alone \citep[e.g.,][]{yon03} 
can be useful for the high-$z$ GRBs. 
In addition, the search for the Ly$\alpha$ break is considered to
be useful. Indeed, the spectroscopic detection of the sharp Ly$\alpha$
cut-off (see Figs.2 and 3) is an accurate method to determine the
redshift. However, the color selection technique which is often used to
find the $z\ga3$ galaxies \citep{mad95,ste96} may not be good for the
high-$z$ GRBs. This is because the GRB afterglows beyond the drop-out
redshift of the considered filter can be detectable if the universe is
ionized enough. 

For example, if we use $I-J \geq 5$ mag as a criterion to select the
high-$z$ GRBs, the selected objects have $z\ga 6.5$ indeed (see Fig.3a
and Table 3). However, a number of real $z\ga 6.5$ objects escape the
criterion if the neutral fraction is less than $\sim 10^{-5}$.
Therefore, we cannot use a simple color selection technique for the
high-$z$ GRBs. Spectroscopic observations to detect the Ly$\alpha$ break
feature , Fe line \citep{meszaros03}  or empirical methods by using only
the $\gamma$-ray data 
\citep{fen00,norris00,in01,ama02,att03,mur03,yon03} are required.

\subsection{Advantage and disadvantage of NIR colors method}

The largest advantage of the NIR color method is its easier
availability and higher sensitivity than other methods. The limiting
magnitude with a high signal-to-noise ratio of the broad-band NIR
photometry reaches much deeper than 20 mag by only a few minute 
exposure with a 8-m class ground based telescope. 
For example, the IRCS (Infrared Camera and Spectrograph)
equipped with the Subaru telescope (Japanese 8-m class 
telescope\footnote{http://subarutelescope.org/}) can reach the 2-$\sigma$
upper limit of 22 mag by only five minutes exposure in $K$-band.
This magnitude corresponds to $\Delta K\simeq 5$ mag based on the
expected $L$ magnitude of $\simeq 16$ mag for the $z\simeq 20$ GRB
afterglows at 1 hour after the burst (Fig.4) and $K-L=1.1$
(Table 3), and also corresponds to the neutral fraction of
$3\times10^{-6}$. Thus, we can put this value as a lower limit of the
neutral fraction at $z\sim 20$ for null detection.

A more strict lower limit can be obtained if we detect the emission from
the reverse shock \citep{sar99a,sar99b,gou03}. Some GRBs show a very
bright early afterglow from the reverse shock. Interestingly, the
apparent magnitude becomes $\sim 5$--6 mag brighter than that shown in
Figure 4. In this case, we can reach $\Delta K\simeq 10$
corresponding to $x_{\rm HI}\simeq6\times10^{-6}$.
A very early observation is highly desired.

{}From the detected magnitude or the limiting magnitude for the null
detection, the neutral fraction or its lower limit in the corresponding
redshift range can be estimated. However, the uncertainty of the
obtained neutral fraction may be large if we have only one photometric
data. This is because the apparent dispersion of the afterglow
luminosities is large. Even in that case, the uncertainty can be
controlled in a low level if we use more than two photometric data,
i.e. colors, which are independent of the absolute luminosity.
This point is one of the important advantages of the NIR color method. 
Therefore, the NIR multi-colors follow-ups of the GRB afterglows 
are strongly encouraged.

\cite{mir98} and \cite{bar03} show that the detailed spectroscopy of
the red damping wing of the Ly$\alpha$ break provides us with the
optical depth for the line, i.e., the neutral hydrogen column density to
the source. However, the spectral resolution required is 
$\lambda/\Delta \lambda \sim 5000(10^5/\tau_{\rm Ly\alpha})$. 
For a lower opacity, a much higher resolving power is needed. 
The limiting magnitude for such observations becomes significantly
shallow. That is, the method of damping wing measurement do not have
sensitivity for a low opacity case. On the other hand, the NIR colors 
method is practically sensitive against $\tau_{\rm Ly\alpha}\la 10$. 
Therefore, these two methods are complementary each other; 
if $x_{\rm HI}\sim10^{-6}$ ($\tau_{\rm Ly\alpha}\sim 1$), the NIR colors
method becomes very useful, whereas the damping wing method is promising
when $x_{\rm HI}\sim 0.1$ ($\tau_{\rm Ly\alpha}\sim10^5$).

Measurement of the CMB polarization anisotropy is also
useful. \cite{hai03} show that we can distinguish how much times the
universe is reionized by using EE spectrum with Planck sensitivity in
3$\sigma$ level. Since the sensitivity of {\it WMAP} is not enough, we
must await the launch of 
Planck\footnote{http://astro.estec.esa.nl/SA-general/Projects/Planck/} 
to reveal the reionization history by the CMB measurement.

Moreover, metal absorption lines like O {\sc i} $\lambda1302$ can be
useful to prove the reionization history \citep{oh02,fur03}. However,
the expected equivalent width is very small as $\la 5$ \AA. We require
the spectroscopy with a resolving power $\sim 5000$. Unfortunately,
observations with a ground-based telescope may be difficult because a
huge number of Earth's atmospheric OH lines conceal the metal lines.
Only JWST (James Webb Space Telescope)\footnote{http://ngst.gsfc.nasa.gov/} 
will have such a high spectral resolution among the future space
telescopes having the NIR spectrograph. Thus, we need to await its
launch to discuss the metal absorption lines.

Technique using the hydrogen 21 cm absorption line is also proposed
(e.g., Madau et al.~1997). However, the brightness of GRB afterglows is
too faint to use them as a background light source for the present and
future radio facilities because the absorption line is very weak
\citep{fur02}.
Thus, we need to look for other sources. Although Carilli et al.~(2002)
proposed luminous high-$z$ radio-loud quasars as a candidate of the
background source, it is very uncertain that they exist at $z\sim 20$.

The dispersion measure in GRB radio afterglows may be promising in near
future \citep{iok03,ino03}. If we observe the radio afterglow at
about 100 MHz within about 1000 s after the burst occurrence, the delay
of the arrival time of the low frequency photons may be detectable by
the Square Kilometer Array. As well as the measurement of the red
damping wing of the Ly$\alpha$ line, this technique is sensitive to
$x_{\rm HI}\ga 0.1$ (i.e., $\tau_{\rm Ly\alpha}\ga 10^5$). Thus, this
technique and our NIR color method are also complementary each other.

The disadvantage of the NIR color method is the coarse redshift
resolution. As discussed in section 4, the neutral fraction obtained by
the color method is an averaged one in the redshift range of 
$z_{\rm Ly\alpha,in} \leq z \leq z_{\rm S}$. Thus, we cannot determine
the so-called reionization redshift. Only the redshift range in which
the reionization occurred is obtained. However, even such a rough
redshift resolution is enough to discuss whether the universe was
reionized once, twice, or more. 

To know the detailed history of the reionization,
the spectroscopy is needed. For this purpose, the Japanese astronomical
satellite ASTRO-F\footnote{http://koala.ir.isas.ac.jp/ASTRO-F/index-e.html} 
can be useful. The satellite will have the spectroscopic sensitivity of
$\sim 30$ $\mu$Jy around $\sim 2$ \micron, which can detect the GRB
afterglows of $z_{\rm S}\sim 20$ if $x_{\rm HI}\sim 10^{-6}$ 
(see Fig.2a). SIRTF (Spece InfraRed Telescope 
Facility)\footnote{http://sirtf.caltech.edu/} may not be useful because
it has the sensitivity only in the wavelength longer than 5 \micron. 
In future, JWST is much promising.
Finally, we note that the narrow-band photometry may be useful because
its narrow transmission width provides us with a moderate resolution of
redshift  keeping a higher sensitivity than that of spectroscopy.

\acknowledgments

We thank Kunihito Ioka for valuable comments.
This work is supported in part by the Grant-in-Aid for Scientific
Research of the Japanese Ministry of Education, Culture, Sports, Science
and Technology, No.14047212 (TN) and No.14204024 (TN), and also
supported by a Grant-in-Aid for the 21st Century COE ''Center for
Diversity and Universality in Physics''. 
AKI and RY thank the Research Fellowships of the Japan Society for the
Promotion of Science for Young Scientists.

\clearpage
\begin{figure}
\plotone{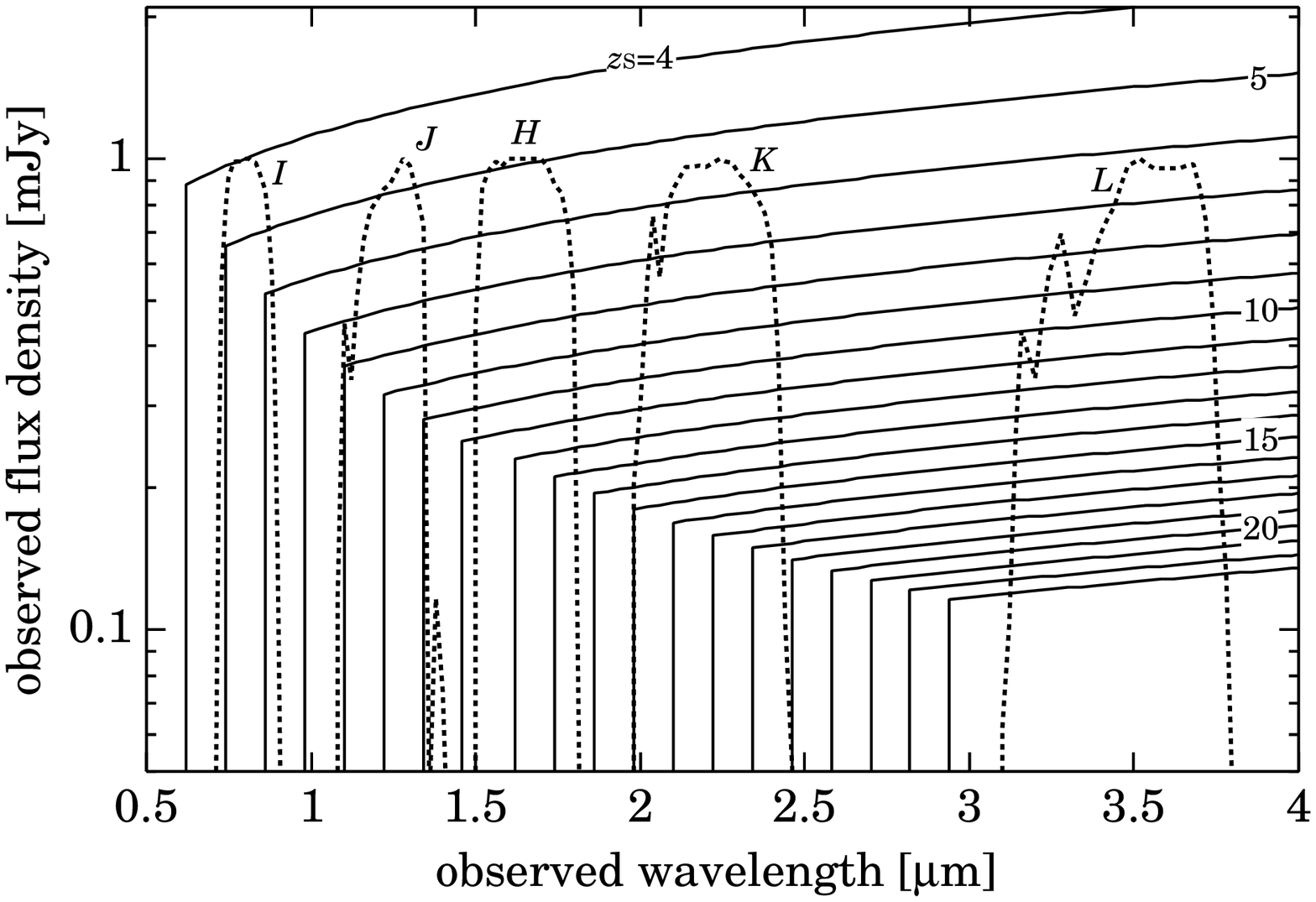}
\caption{Afterglow spectra in the hypothetical neutral universe. The
 solid curves are the afterglow spectra of the gamma-ray bursts at the
 redshift $z_{\rm S}=4$--23 as indicated in the panel. 
 The observing time is set to be 1 hour after
 the burst for all curves. The dotted curves are the filter
 transmissions of $I$, $J$, $H$, $K$, and $L$ bands \citep{bes88,bes90}.
 These transmission curves are scale free.}
\end{figure}

\begin{figure}
\plottwo{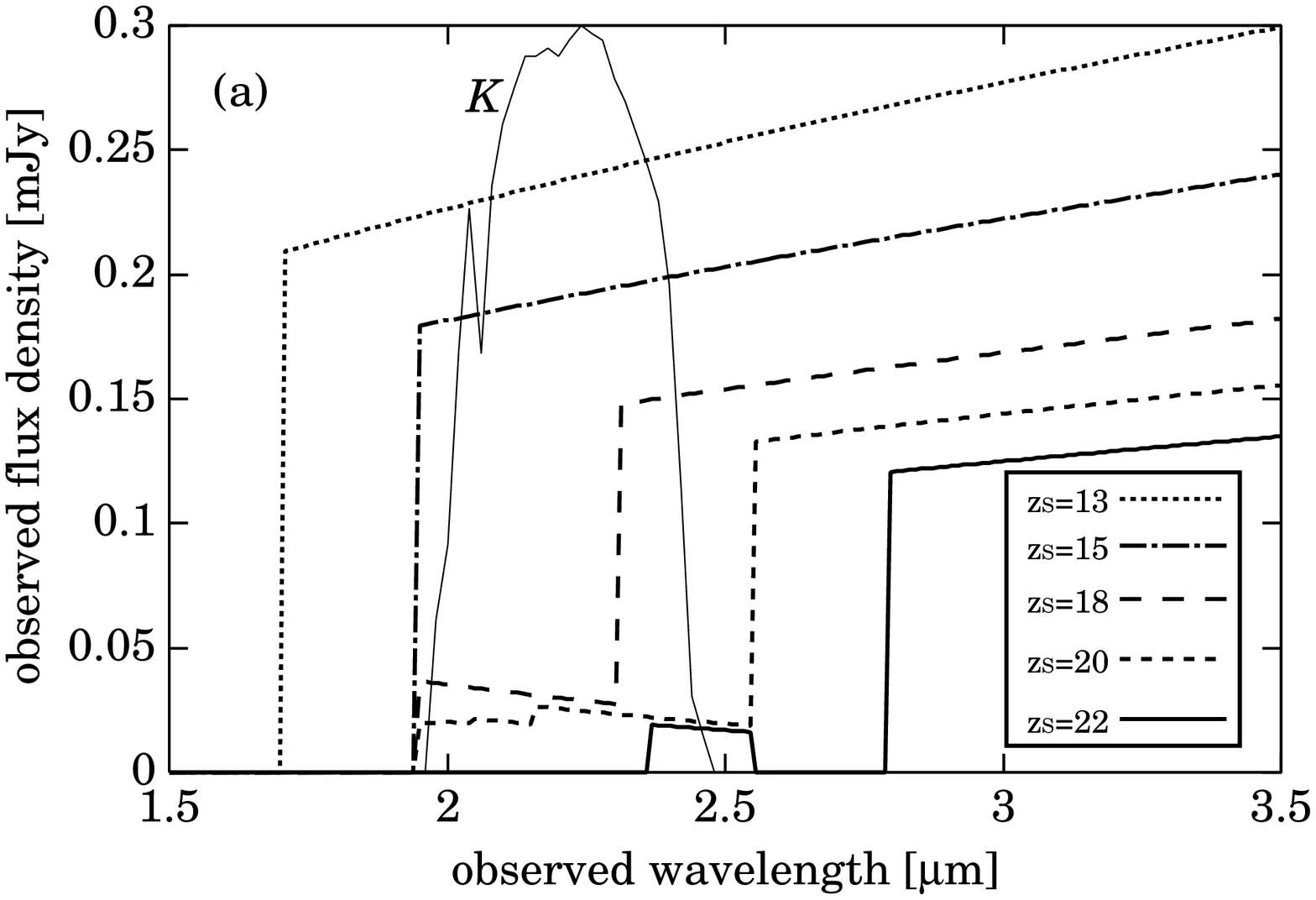}{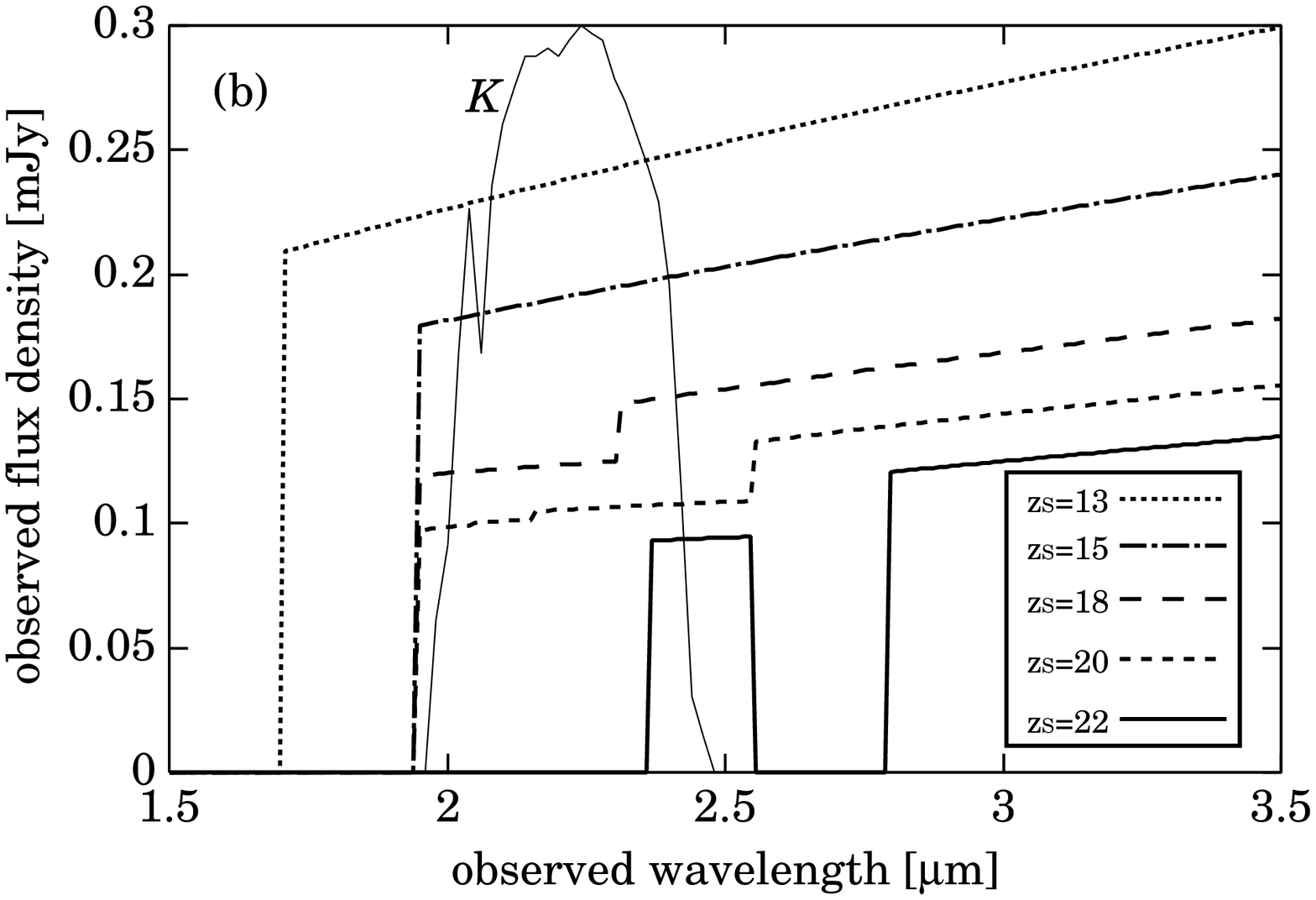}
\caption{Expected near-infrared afterglow spectra of very high redshift
 gamma-ray bursts. The solid, short-dashed, long-dashed, dot-dashed,
 and dotted curves are the cases of the source redshift $z_{\rm S}=22$,
 20, 18, 15, and 13 respectively. The observing time is set to be 1 hour
 after the burst for all curves. The neutral fraction in 
 $15 \leq z < 20$ is assumed to be $10^{-6}$ and $10^{-7}$ for the
 panels (a) and (b), respectively. The thin solid curve is the
 filter transmission of $K$-band \citep{bes88}. 
 This transmission curve is scale free.}
\end{figure}

\begin{figure}
\plottwo{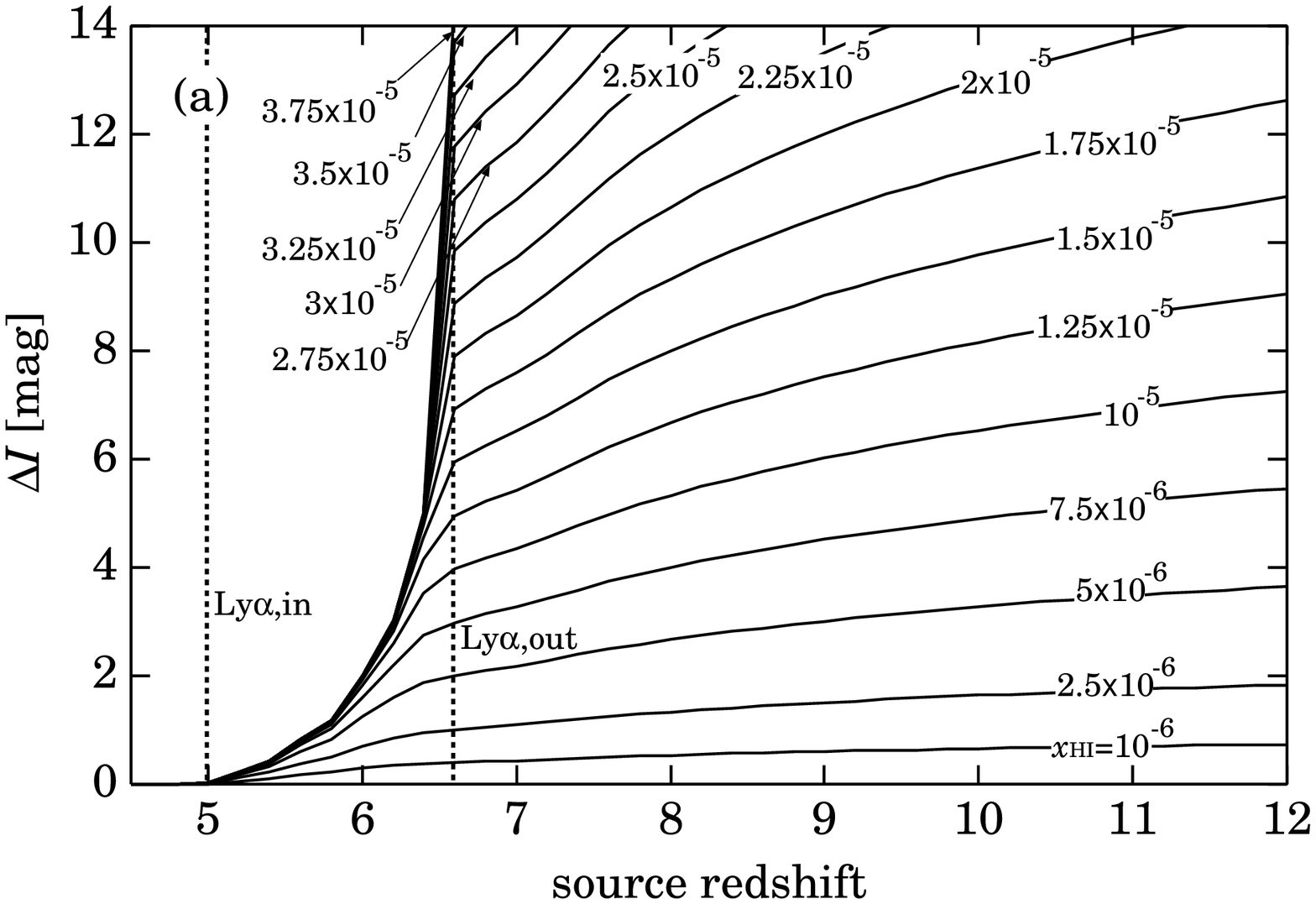}{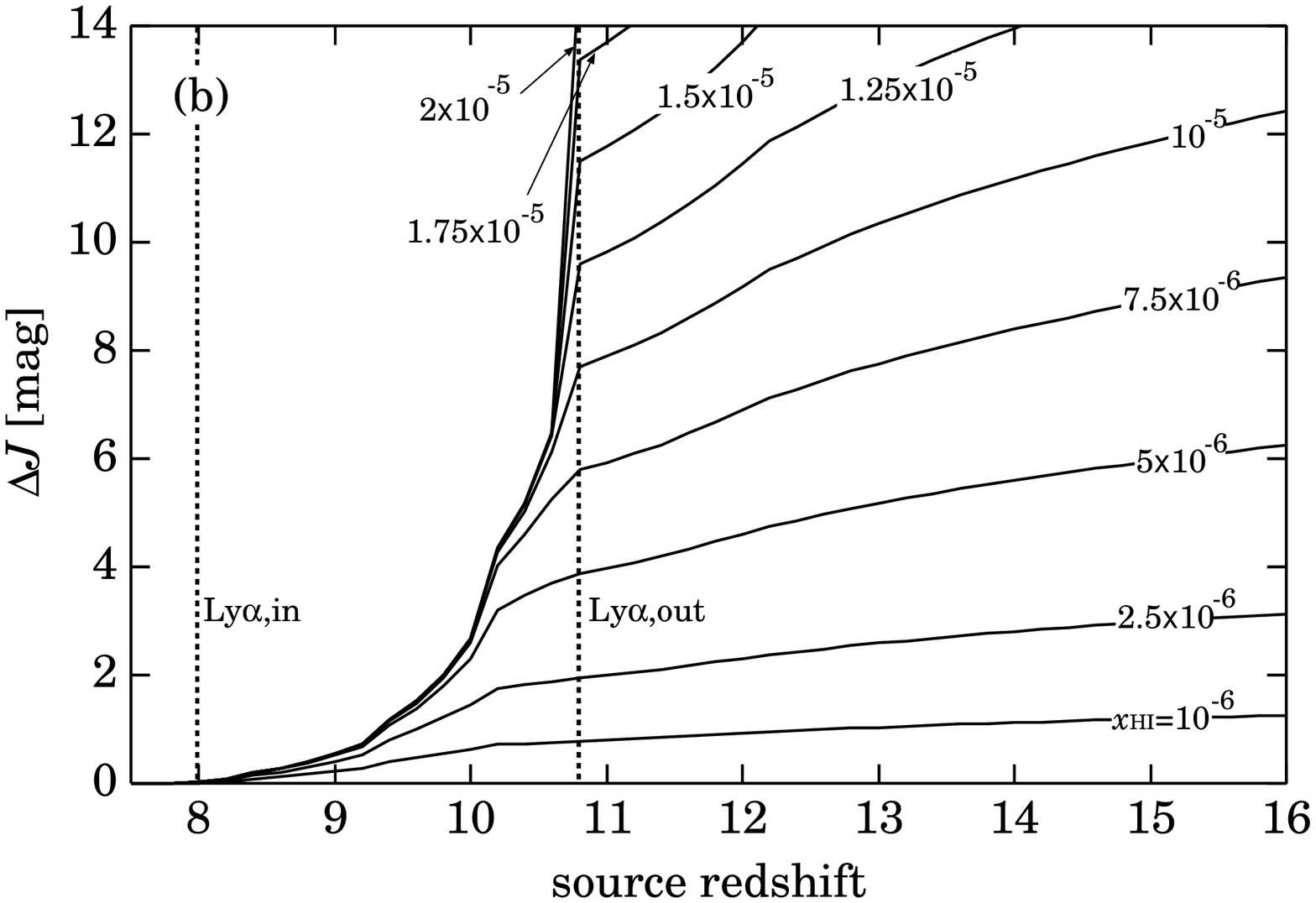}
\plotone{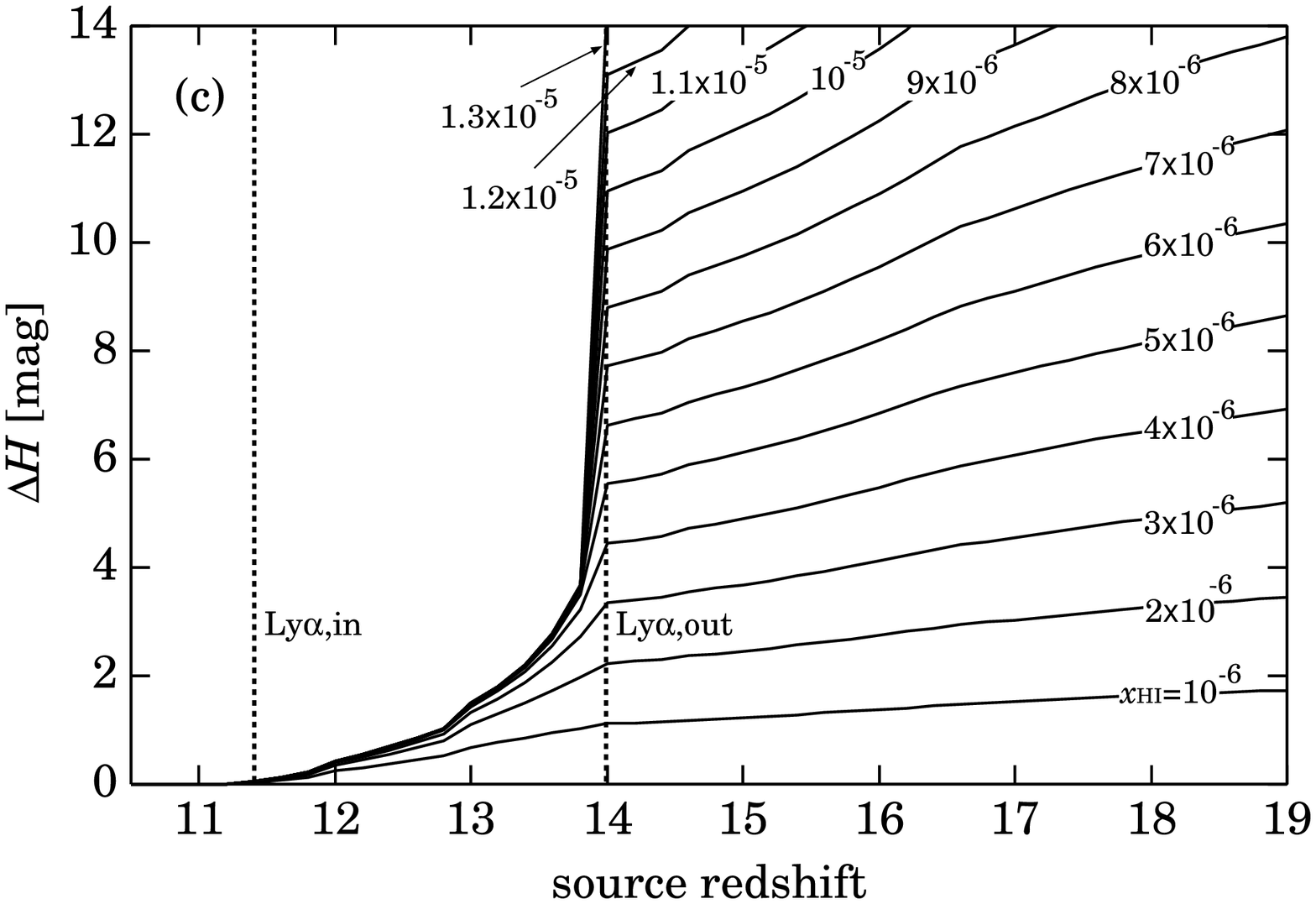}
\plotone{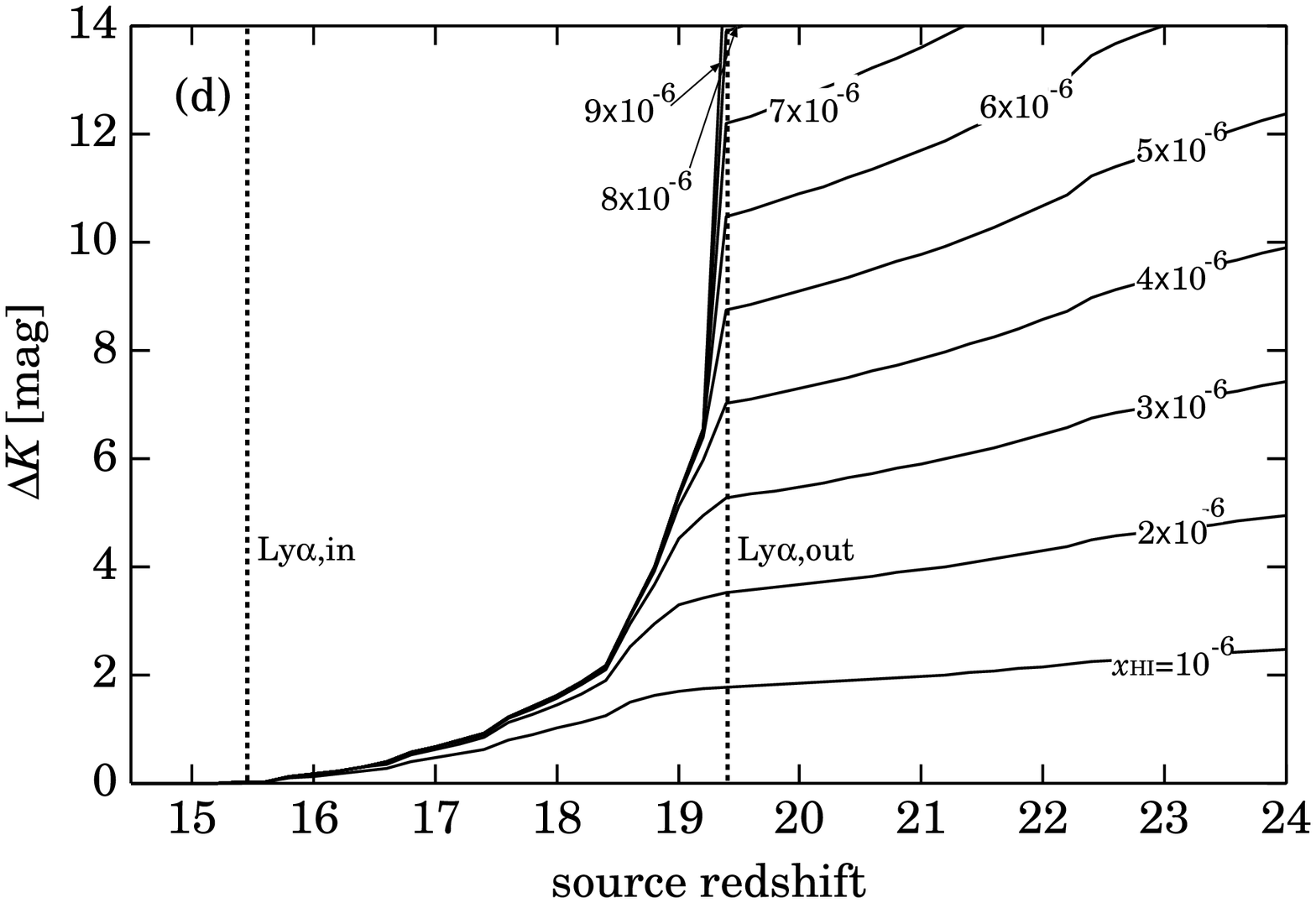}
\caption{The intergalactic absorption in near-infrared bands. The
 vertical axis in each panel means the absorption amount in each filter
 (see eq.[11]); (a) $I$-band, (b) $J$-band, (c) $H$-band, and (d)
 $K$-band. The solid curves in the panels are loci of the absorption
 amount as a function of the redshift of the source $z_{\rm S}$. The
 neutral hydrogen fraction, $x_{\rm HI}$, is assumed to be constant
 between the redshift when the Ly$\alpha$ break enters the filter
 transmission and $z_{\rm S}$. The assumed $x_{\rm HI}$ is indicated on
 each curve. Two dotted vertical straight lines in each panel
 indicate the source redshifts at which the Ly$\alpha$ break enters and
 goes out of the filter.}
\end{figure}

\begin{figure}
\plotone{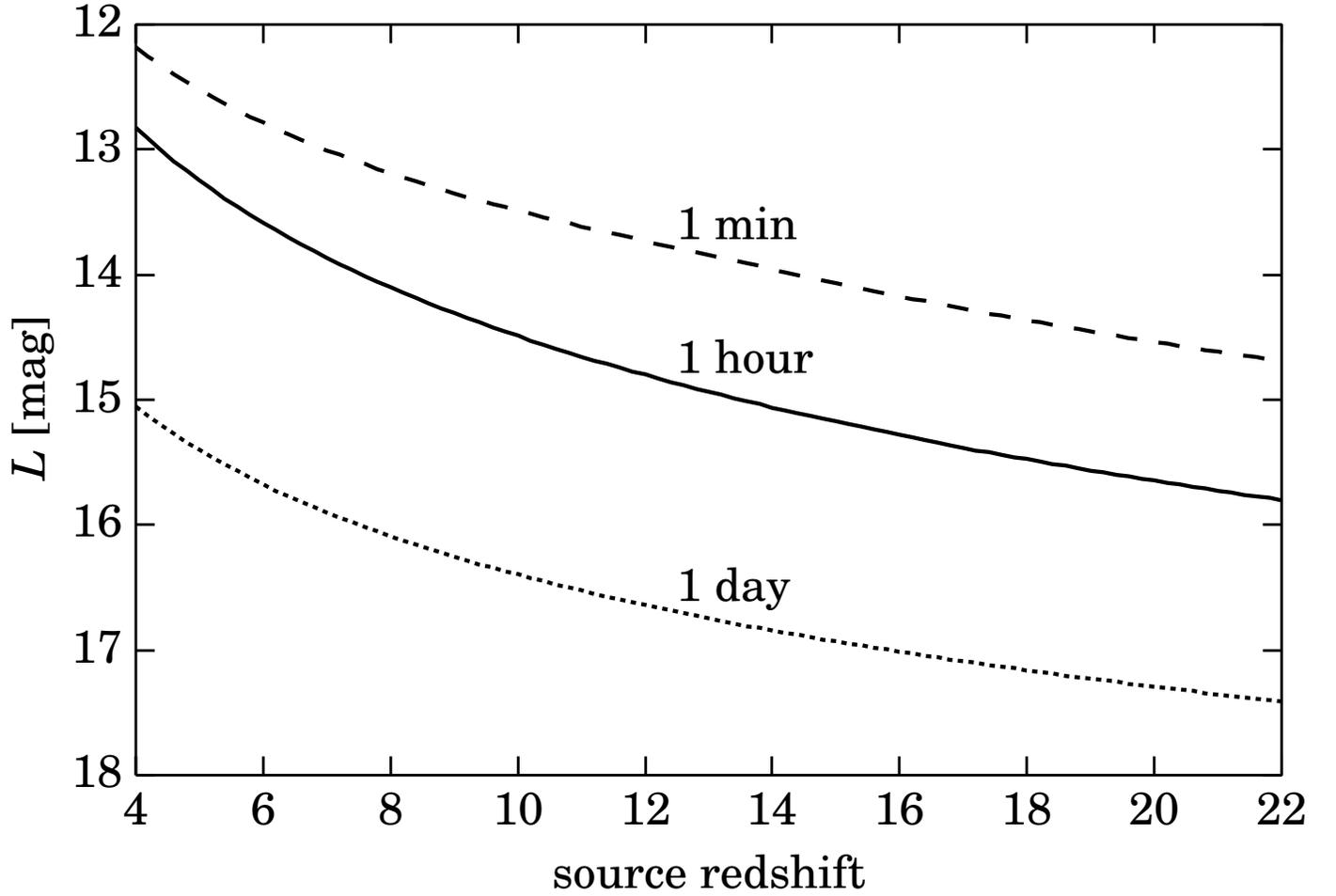}
\caption{The expected apparent magnitude in $L$-band of the afterglows
 of the gamma-ray bursts (GRBs) as a function of the source
 redshift. The dashed, solid, and dotted curves are the cases
 of 1 minute, 1 hour, and 1 day after the burst occurrence in the
 observer's frame, respectively. The spectral model of the GRB
 afterglows by \cite{ciardi00} is adopted.  The assumed parameter set is
 described in the first paragraph of section 3.}
\end{figure}

\begin{figure}
\plotone{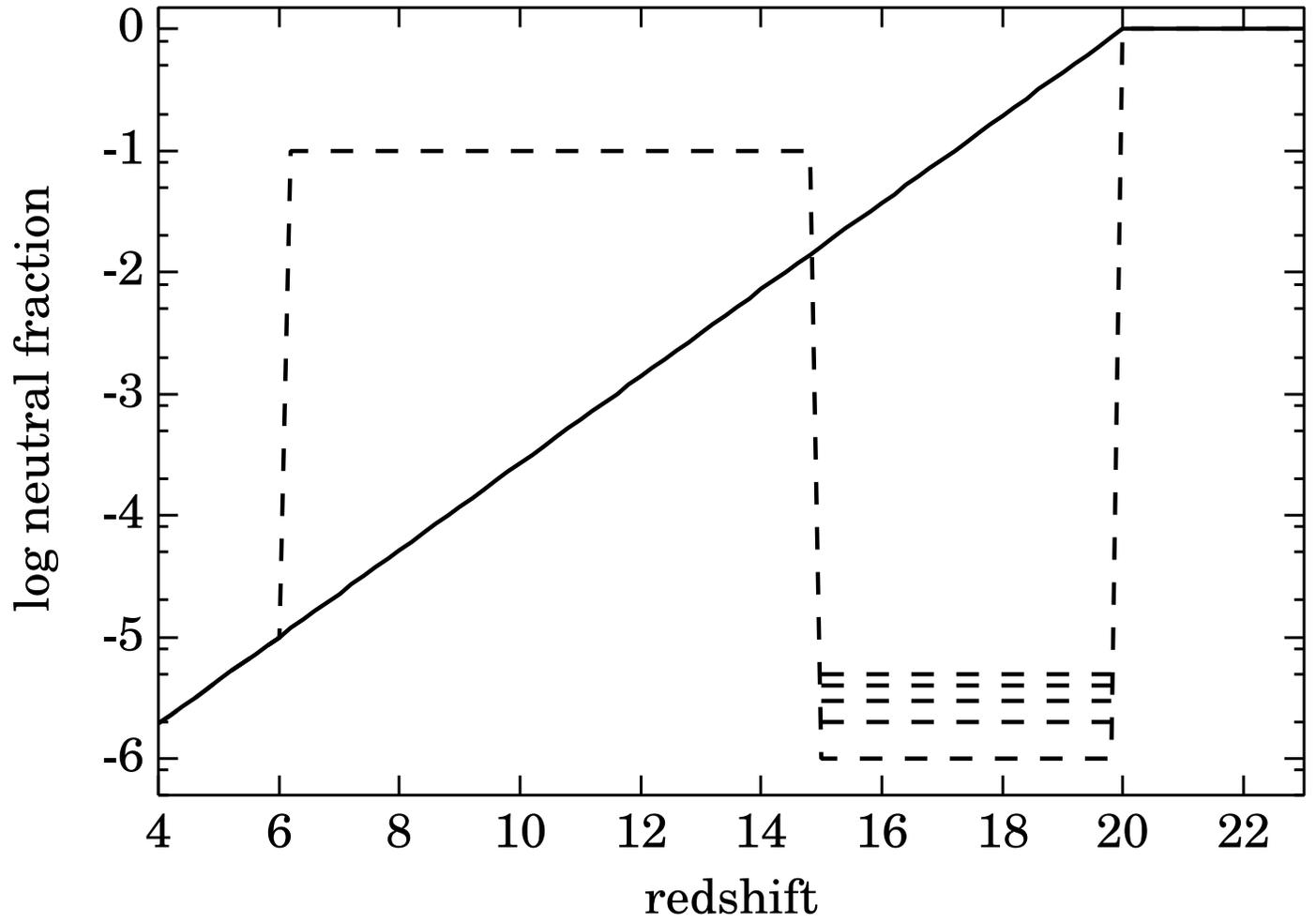}
\caption{Examples of the reionization history. The solid curve is a
 single gradual reionization case. The dashed curve is a double
 reionizations case as suggested by \citet{cen03a}.}
\end{figure}

\begin{figure}
\plottwo{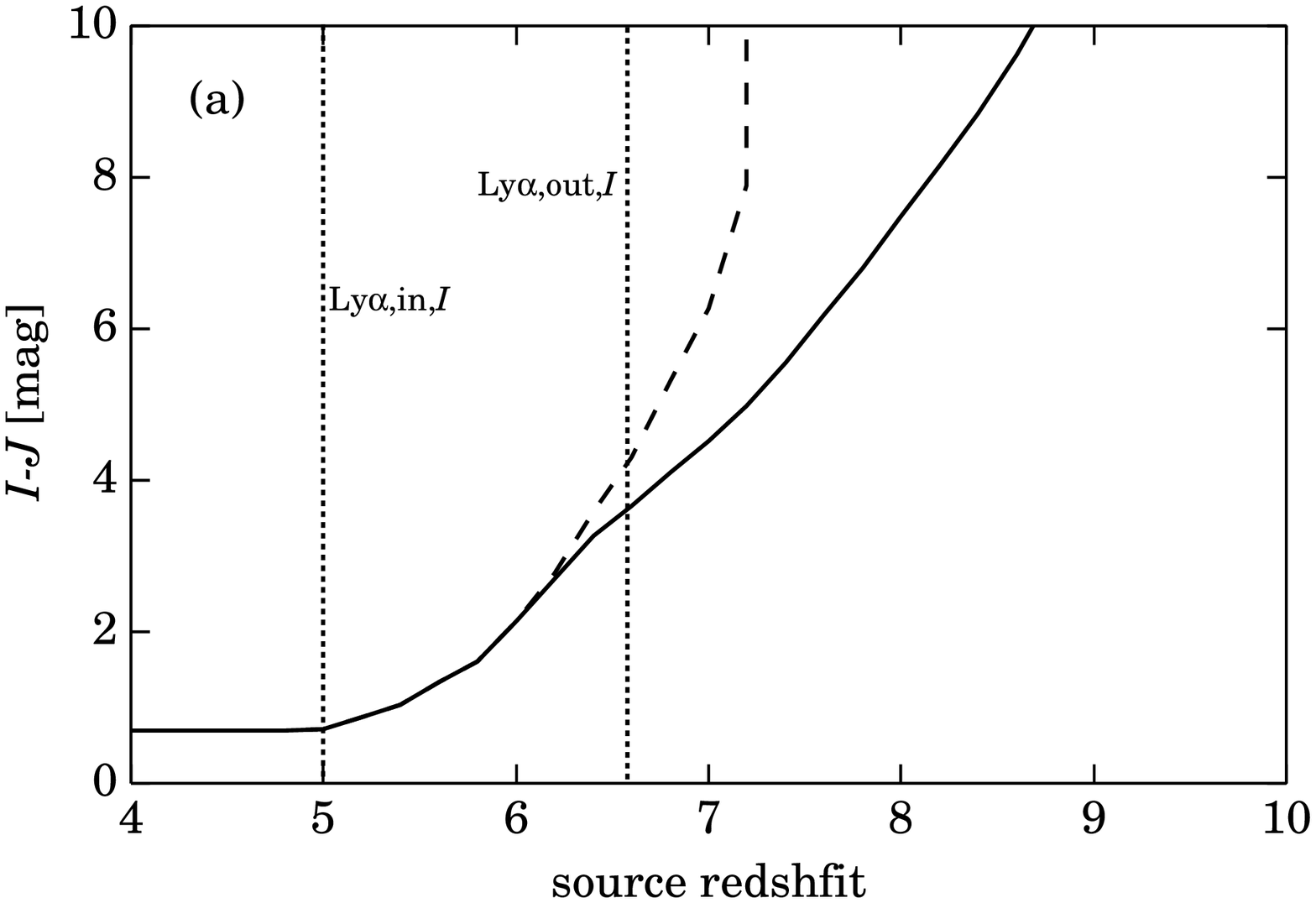}{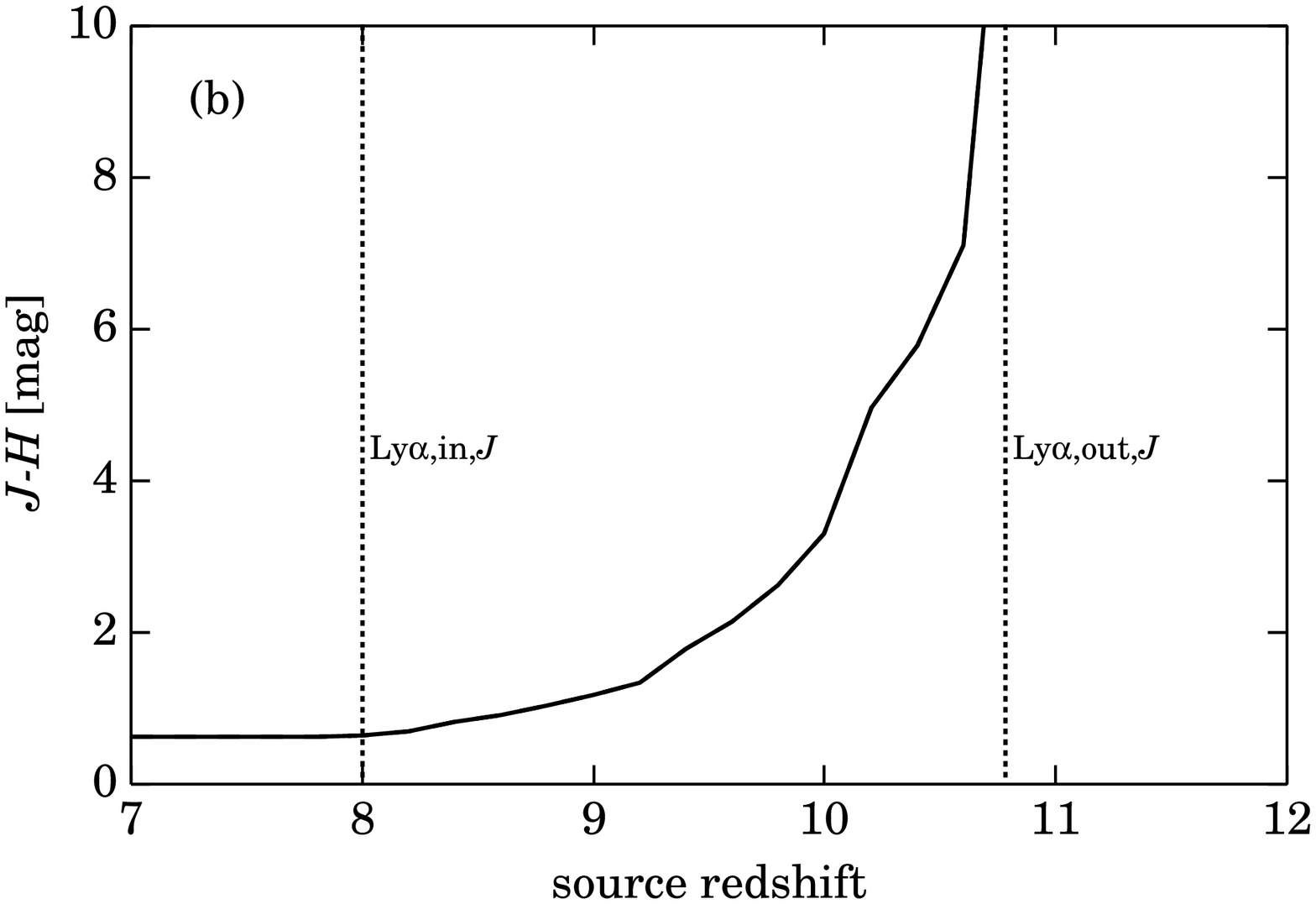}
\plotone{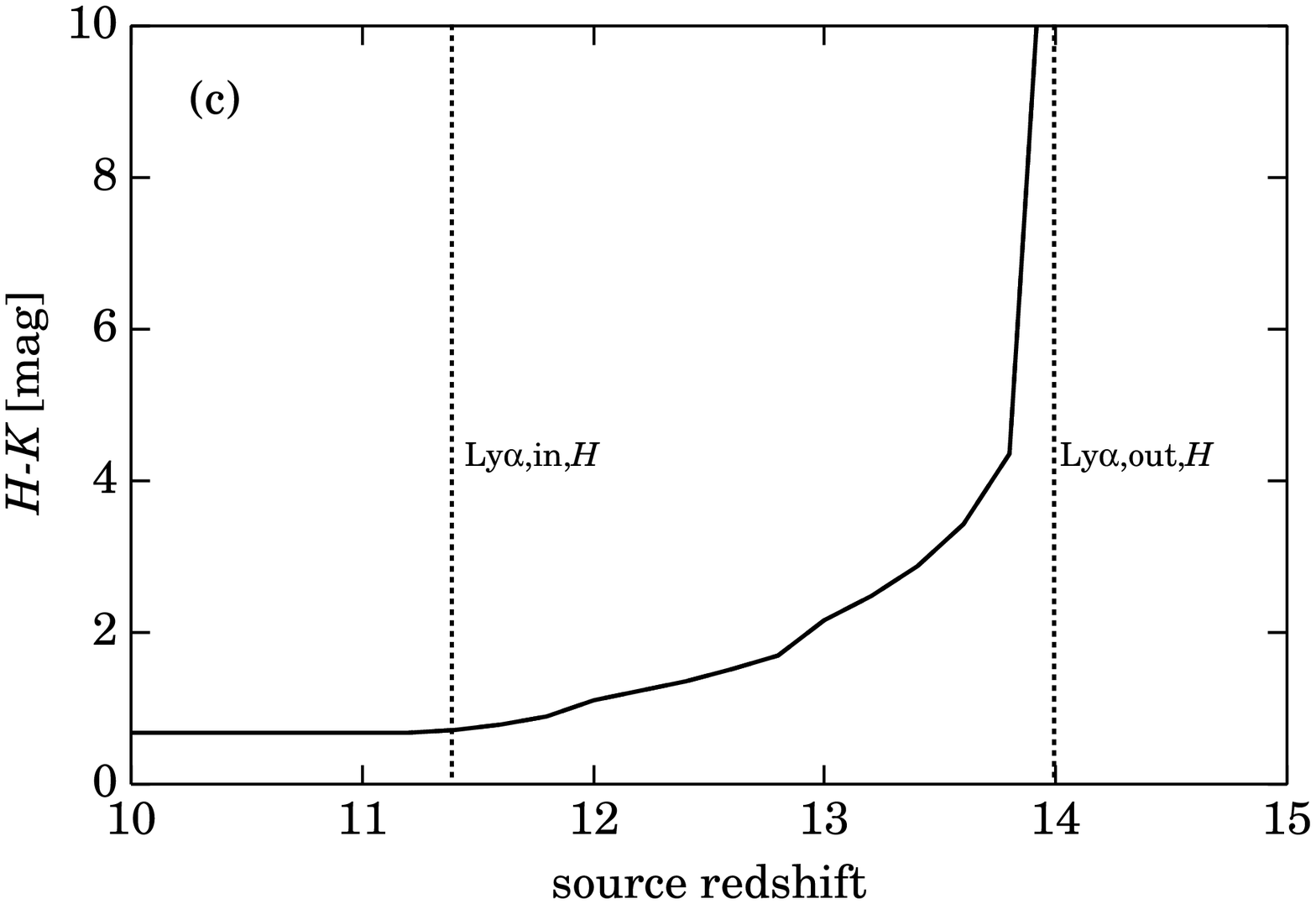}
\plotone{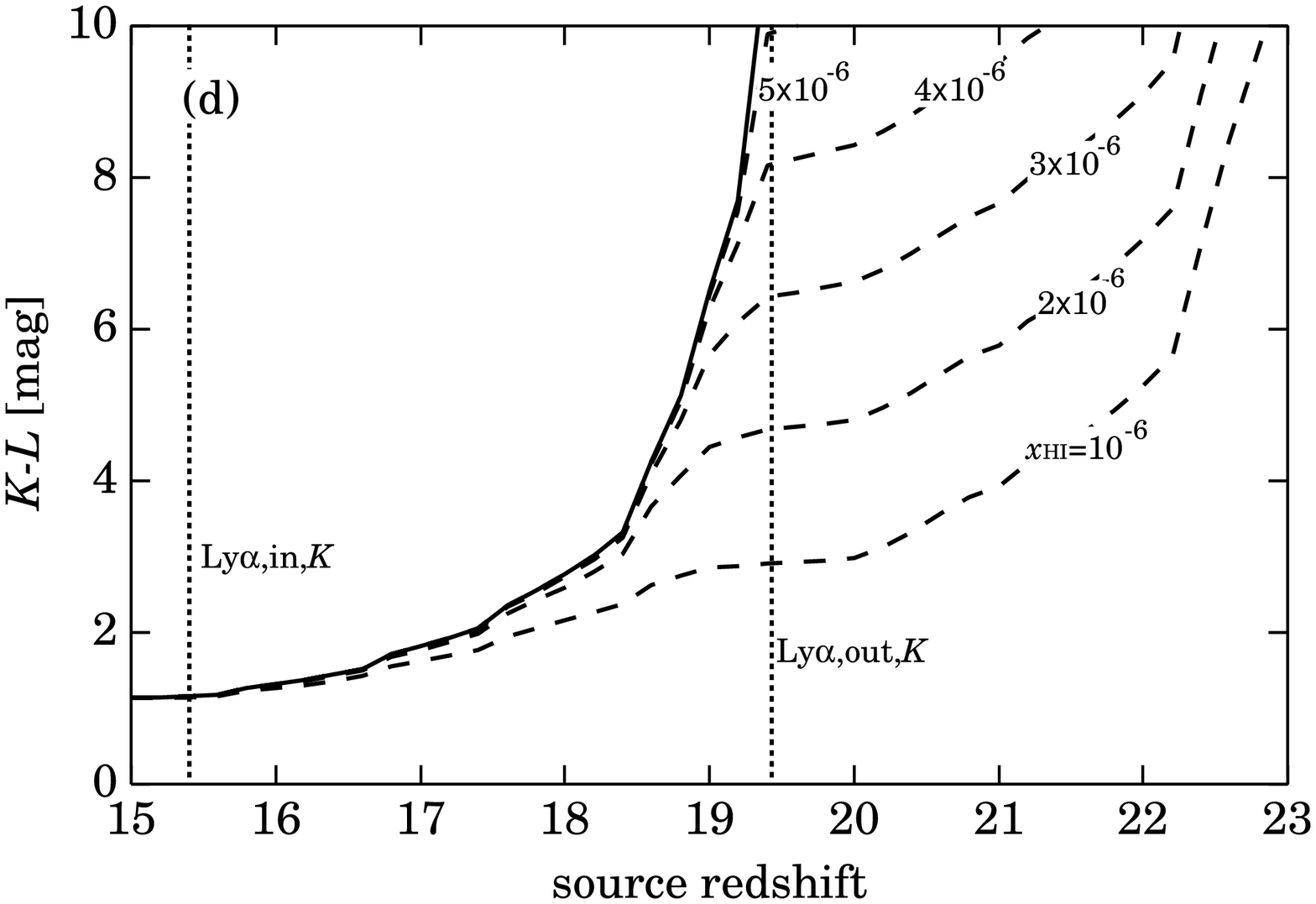}
\caption{Expected near-infrared colors of gamma-ray burst
 afterglows. (a) $I-J$ colors, (b) $J-H$ colors, (c) $H-K$ colors, and
 (d) $K-L$ colors. The solid and dashed curves correspond to the cases
 of the single and double reionizations depicted in Figure 5,
 respectively. The solid and dashed curves in panels (b) and (c) are
 perfectly superposed. In panel (d), some cases of the neutral hydrogen
 fraction in the first reionization are shown. Two dotted vertical
 straight lines in each panel indicate the source redshifts at which the
 Ly$\alpha$ break enters and goes out of the indicated filter.}
\end{figure}

\clearpage

\begin{deluxetable}{ccc}
\tabletypesize{\scriptsize}
\tablecaption{Star formation density required to maintain 
$x_{\rm HI}\sim10^{-6}$}
\tablewidth{0pt}
\tablehead{
\colhead{Stellar mass range} & \colhead{$\epsilon_{\rm LC}$} 
& \colhead{$\rho_{\rm SFR}$}\\
\colhead{($M_\sun$)} & \colhead{($10^{61}$ ph ${M_\sun}^{-1}$)} 
& \colhead{($M_\sun$ yr$^{-1}$ Mpc$^{-3}$)}
}
\startdata
10--100 & 2 & 0.05 \\
1--100 & 0.7 & 0.1 \\
0.1--100 & 0.3 & 0.3 \\
\enddata

\tablecomments{We calculate the Lyman continuum photon emissivity per
 unit stellar mass, $\epsilon_{\rm LC}$, by Starburst 99 model
 \citep{lei99} by assuming the Salpeter initial mass function with the
 tabulated mass range. We also assume the escape fraction 
 $f_{\rm esc}=0.1$ and the clumping factor $C=1$ for the star formation
 density.}

\end{deluxetable}

\begin{deluxetable}{ccc}
\tabletypesize{\scriptsize}
\tablecaption{Characteristic redshifts of NIR filters}
\tablewidth{0pt}
\tablehead{
\colhead{filter} & \colhead{$z_{\rm Ly\alpha,in}$} 
& \colhead{$z_{\rm Ly\alpha,out}$}
}
\startdata
$I$ & 5.0 & 6.6 \\
$J$ & 8.0 & 10.8 \\
$H$ & 11.4 & 14.0 \\
$K$ & 15.4 & 19.4 \\
$L$ & 24.6 & 30.6 \\
\enddata

\tablecomments{$z_{\rm Ly\alpha,in}$ and $z_{\rm Ly\alpha,out}$ are the
 redshifts at which the Ly$\alpha$ break enters and goes out of the
 filter transmission, respectively.}

\end{deluxetable}

\begin{deluxetable}{ccc}
\tabletypesize{\scriptsize}
\tablecaption{NIR afterglow intrinsic colors}
\tablewidth{0pt}
\tablehead{
\colhead{color} & \colhead{$\propto \nu^{-1/2}$} 
& \colhead{$\propto \nu^{-p/2}$ ($p=2.5$)}
}
\startdata
$I-J$ & 0.70 & 1.0\\
$I-H$ & 1.3 & 1.9\\
$I-K$ & 2.0 & 2.8\\
$I-L$ & 3.1 & 4.3\\
$J-H$ & 0.62 & 0.86\\
$J-K$ & 1.3 & 2.8\\
$J-L$ & 2.4 & 3.3\\
$H-K$ & 0.67 & 0.91\\
$H-L$ & 1.8 & 2.4\\
$K-L$ & 1.1 & 1.5\\
\enddata

\tablecomments{In the observer's frame, the NIR afterglow spectrum of
 the GRBs at $z\sim 10$ is proportional to $\nu^{-1/2}$ between $\sim 1$
 minute and several hours after the initial burst and proportional to
 $\nu^{-p/2}$ for later time.}

\end{deluxetable}

\end{document}